\def\e{\mbox{e}}
\def\d{\mbox{d}}
\documentstyle[aps,pre]{revtex}
\begin{document}
\draft
\title{ Protein adsorption on lipid  monolayers at their 
        coexistence region
}
\author{
Roland R. Netz\cite{address} and David Andelman}

\address{
School of Physics and Astronomy, 
Raymond and Beverly Sackler Faculty of Exact Sciences,\\
Tel Aviv University,
Ramat Aviv 69978, Tel Aviv,  Israel}

\author{H. Orland\cite{address2}}
\address{
Service de Physique Theorique, CE-Saclay, 91191 Gif sur Yvette, Cedex, France}
\date{\today}
\maketitle

\begin{abstract}
We investigate theoretically the behavior of proteins
as well as other large macromolecules which are incorporated into
amphiphilic monolayers at the air-water interface.
We assume the monolayer to be in the coexistence
region of the ``main'' transition, where
domains of the liquid condensed phase coexist with
the liquid expanded background.
Using a simple mean-field free energy accounting
for the
interactions between proteins and amphiphilic molecules, we obtain 
the spatial protein distribution with the following
characteristics.
When the proteins preferentially interact with either the 
liquid condensed or liquid expanded domains, they
will be dissolved in the respective phase.
When the proteins are energetically rather indifferent
to the density of the amphiphiles,
they will be localized at the line boundary
between the (two-dimensional) liquid expanded and condensed phases.
In between these two limiting cases, a delocalization transition
of the proteins takes place. This transition
 is accessible by changing the temperature
or the amount of incorporated protein.
These findings are in agreement with recent fluorescence 
microscopy experiments.
Our results also apply to lipid multicomponent membranes
showing coexistence of distinct fluid phases.
\end{abstract}

\pacs{82.70, 87.20E, 64.75}

\section{Introduction}

Monolayers of amphiphilic molecules spread on liquid surfaces have
traditionally been studied as models for biological 
membranes\cite{Gaines,Bloom}.
Such insoluble and monomolecular films made of suitable phospholipids 
or fatty acids are stable over a wide
range of surface pressures and temperatures due to the strong 
reduction of the water surface tension and are called 
{\it Langmuir monolayers}\cite{Mohwald}.
In typical experiments, the amphiphiles are
solubilized in a volatile solvent and placed
on the air-water interface. As the solvent
evaporates, the amphiphiles  spontaneously spread and form a monolayer.
When the insoluble film is then compressed (while keeping the temperature
fixed), the lateral pressure can be measured
as a function of the area per amphiphilic  molecule in  
analogy to bulk isotherms.

Using film balance
techniques\cite{Gaines,Bloom,Mohwald,Albrecht1,Albrecht2},
the following general picture emerged.
When the extremely expanded film is compressed,
it produces the liquid expanded phase (LE), which, at low
enough temperatures, transforms upon further compression 
into the liquid condensed phase (LC).
At much lower surface concentrations and at low enough temperatures,
the monolayer undergoes a first-order transition into a
gaseous phase. At very high lateral pressures, solidification occurs,
as indicated by a discontinuity in the pressure-area isotherms.
Subsequently, these systems were also 
studied using X-ray\cite{Mohwald2,Dutta,Rondelez,Schlossmann} and
neutron\cite{neutron} scattering techniques,
indicating the existence of a large number of different condensed phases.
In this paper we will be concerned only with the LE/LC transition.
Therefore, we do not introduce appropriate order parameters
needed to distinguish the different condensed phases\cite{Bibo}.

The nature of the LE/LC transition 
has been the subject of much discussion\cite{Pallas}.
It is analogous to the ``main'' transition in 
lipid bilayers\cite{Mohwald},
where domains of the LE and the LC phase have
been shown to coexist over a wide range of lipid surface concentrations
(or area per molecule).
In this coexistence region, the condensed domains show a large
variety of different shapes\cite{McConnell} 
and grow as the area per molecule is decreased,
whereas the number of domains depends on the initial conditions
and typically stays fixed.
The isotherms in the coexistence region, however, were found to be
non-horizontal, which led to the  postulation of a limited
cooperativity of this transition\cite{Albrecht1}.
For the case of single-chain fatty acids,
it was later shown  that the isotherms  approach
zero slope as the material used is progressively purified\cite{Pallas}.

On the theoretical side, the LE/LC transition has been modeled based
on various microscopic pictures of the interaction between
surfactant (or lipid) molecules including translational as well as 
internal degrees of freedom\cite{Marcelja,Bell,Georgallas,Chen,Kramer}.

The biological function of membranes depends mostly on the
incorporation of proteins and other macromolecules into 
the lipid layers. Functionality and efficiency of these inclusions
depend crucially on microscopic details of the embedding in the
lipid matrix,
 which can occur in different
ways. Monolayers at the air-water interface are suitable for
the study of the interaction between lipids and proteins,
since they are rather well-defined and allow
the control of independent thermodynamic parameters which
are otherwise fixed in a bilayer membrane, like the area per molecule.
Also, the observational techniques are well developed.
Direct visualization  of the phase behavior of monolayers
can be obtained using fluorescence microscopy techniques.
Here, a fluorescent dye probe is incorporated
into the monolayer the lateral distribution of which can be obtained
from the analysis of fluorescence micrographs. 
Contrast in the images is obtained as a result of different dye
solubility, fluorescence quantum yield, or molecular density
of coexisting phases\cite{Fluor}.
A complementary and recently developed technique
 is Brewster-angle microscopy, which allows
imaging of a monolayer without the addition of fluorescent 
probes\cite{Henon}.

After injection of a water-soluble protein into the aqueous subphase,
the surface tension typically decreases, indicating that
the protein is at least partially incorporated into
the monolayer\cite{Mohwald,Wiedmann,Schwinn,Vogel}.
This is due to the protein affinity  to the water/air interface.
The specific type of this attraction is not well understood  and
probably is due in part to
structural changes (denaturation) of the protein in the
monolayer or at the water  surface, associated with the unfolding 
of hydrophobic groups.

One of the striking experimental
observations\cite{Schwinn,Haas} was that some proteins
adsorb preferentially along the {\it boundary line} between the
LE and LC domains  when the monolayer is in the LE/LC coexistence region. 
These observations were made for
fluorescently labeled small proteins,
such as {\em concanavalin} A\cite{Haas}
or {\em streptavidin}\cite{Schwinn},
interacting with phospholipid monolayers.
These experimental
findings motivated our present theoretical study.

In the following, we describe a simple
model, which (i) assumes 
the LE/LC transition to be a simple first-order condensation
transition, yielding  coexisting domains for
temperatures below the critical temperature, 
and  (ii) includes the effect of proteins which
are adsorbed into the monolayer. Assuming that the proteins
are completely incorporated into the monolayer, this 
simplistic model leads to an entropic force which tends to
localize the protein at the boundary between LE and LC domains.
Depending on the energetic
preference of the protein for the LE or LC phase, the protein
will be either dissolved in the LE or in the LC domain, or,
if there is no pronounced preference, will be
localized at the boundary.

Phase separation in amphiphilic layers is  also observed
for freely suspended multicomponent bilayers\cite{Bloom}. Here,
the coexisting phases are distinguished by their compositions.
The most important examples include mixtures of phospholipids
with cholesterol\cite{Thewalt} and mixtures of different phospholipids\cite{Wu},
and in both cases the coexisting phases are in a fluid state.
These phenomena are of great biological interest since biological membranes
are always multicomponent mixtures  and lateral organization into domains
is supposed to play an important functional role. We note that our results apply
directly to these situations as well, although we will limit our terminology
to the situation of coexisting dense and dilute phases for one-component systems
at the air-water interface. For the case of freely suspended
membranes,  our findings  imply a simple 
mechanism for the localization  of integral membrane  proteins 
along the one-dimensional boundary between coexisting domains.
The resulting enrichment  of proteins might be a prerequisite for proper
biological function in certain cases. 

In the following sections we formulate the model (Sect. II),
inspect the minima of the free energy (Sect. III), 
solve the corresponding Euler-Lagrange equations
in the coexistence region (Sect. IV),
and calculate profiles both for the lipid and the (coupled)
protein densities (Sect. V).
From the profiles we generate a general phase diagram 
featuring localized, semi-localized and delocalized
protein phases.
We also calculate the total amount of adsorbed protein,
the protein excess $\Gamma$ (Sect. VI), and the line tension 
$\tau$ of the LE-LC line interface (Sect. VII). 
It turns out that the line tension is strongly
reduced by the adsorption of proteins. A finite solubility 
of the proteins in the subphase is taken into account in Sect. VIII.
Finally, the connection to experimentally measurable 
quantities, such as the surface pressure $\Pi$, is made
in Sect. IX.

\section{The mixed lipid and protein  free energy}
Consider the  air-water interface with proteins, lipid molecules,
and artificial ``vacancies'', with area fractions
$\phi_P$, $\phi_L$, and $\phi_V$, respectively,
satisfying $\phi_P+\phi_L+\phi_V=1$.
The vacancies are introduced in order to allow for independent variations
of the protein and lipid concentrations,
hence making  coexistence of dilute and condensed regions of the
monolayer possible.
Inscribing the system on a lattice,
with a lattice constant corresponding to the size of a lipid molecule,
the free energy of mixing per lattice site  within a mean 
field theory can be written
for the three-component mixture as a sum of the 
enthalpy and entropy of mixing,
${\cal F} = {\cal U} -T{\cal S}$.
The enthalpy of mixing includes all pair-wise
interactions between the three species:
\begin{equation}
{\cal U}/T=
E_{LL} \phi_L^2 +E_{VV} \phi_V^2 +E_{PP} \phi_P^2
+E_{LV} \phi_L \phi_V +E_{PL} \phi_P \phi_L+E_{PV} \phi_P \phi_V
\end{equation}
and the $E_{ij}$ are the dimensionless interaction parameters for all
possible pairs.
The entropy of mixing is
related to the total number 
$\Omega$ of distinct microscopic configurations
\begin{equation}
{\cal S}=\frac{\log \Omega}{N}
\end{equation}
where $N$ is the total number of lattice sites and the Boltzmann
constant is set to unity ($k_B=1$).
In the random-mixing approximation,
\begin{equation}
\Omega=
\frac{(N/\alpha)!}{(N \phi_P /\alpha)! (N(1-\phi_P)/\alpha)!}
\frac{(N[\phi_L+\phi_V])!}{(N \phi_L)!(N \phi_V)!}
\end{equation}
where the constant $\alpha>1$  denotes the ratio between the compact 
area occupied by a protein molecule and a lipid molecule at the interface.
The above expression is the product of the number of all protein  configurations
and the number of all lipid/vacancy configurations in the remaining area
not taken up by the proteins.
Using Stirling's Formula in the thermodynamic limit, defined by
 $N \rightarrow \infty$, the expression for ${\cal S}$ can be simplified 
\begin{equation}
{\cal S}=-\phi_L \log(\phi_L) -\phi_V \log(\phi_V) -\phi_P \log(\phi_P)/\alpha
-(1/\alpha -1)(1-\phi_P) \log (1-\phi_P)
\end{equation}
It is convenient to define the thermodynamic potential
\begin{equation}
{\cal G}/T=
{\cal F}/T- \mu_P \phi_P -\mu_L(\phi_L-\phi_V)
\end{equation}
where the chemical potentials $\mu_P$ and $\mu_L$  are 
coupled to 
the protein concentration $\phi_P$ and the difference
between the lipid and vacancy concentrations, $\phi_L-\phi_V$,
respectively.

In (1)-(5),
long-range interactions between the proteins, 
such as electrostatic forces, are not taken into account. 
In addition, the free energy of mixing assumes a confinement 
of the protein and lipid
to the two-dimensional plane of the air-water interface.
In fact, the variation of the protein concentration
perpendicular to the monolayer in the subphase can be taken 
into account approximately and
leads to a renormalization of the parameters
of the two-dimensional model, as shown in Sect. VIII.

The lipid order parameter $\eta$, corresponding to the
density of lipid molecules, can be written  as
\begin{equation}
\eta \equiv \phi_L - \phi_V
\end{equation}
Using  that $\phi_P + \phi_L +\phi_V=1$, 
and defining the protein concentration
as $\phi \equiv \phi_P$,
the free energy ${\cal F}$ and the potential ${\cal G}$  can be rewritten as 
\begin{eqnarray}
{\cal F}/T &=&
-(J+1/2) \eta^2 +L \phi^2
+ \lambda \eta \phi+ \\ \nonumber
&& (1+\eta-\phi)\log[(1+\eta-\phi)/2]/2+
   (1-\eta-\phi)\log[(1-\eta-\phi)/2]/2+\\ \nonumber
&& \phi \log[\phi]/\alpha +(1/\alpha-1) (1-\phi) \log[1-\phi]
\end{eqnarray}
and 
\begin{equation}
{\cal G}/T= {\cal F}/T- \mu_{\eta} \eta-(\mu +\log 2 ) \phi
\end{equation}
where constant terms have been omitted and
 linear terms in $\eta$ and $\phi$ have been dropped out from
${\cal F}$  for convenience.
They merely contribute a constant shift to $\mu$ and
$\mu_{\eta}$ in ${\cal G}$.
The reduced interaction parameters: 
$J$, $L$, $\mu$ and $\lambda$ are related to the 
original $E_{ij}$ and $\mu_P$ in the following way
\begin{equation}
-J \equiv 
\frac{1}{4}(E_{LL} + E_{VV} - E_{LV})+\frac{1}{2}
\end{equation}
\begin{equation}
L\equiv \frac{1}{4}(E_{LL} + E_{VV} +  E_{LV})+ 
E_{PP}-\frac{1}{2}(E_{PL} + E_{PV})
\end{equation}
\begin{equation}
 \lambda \equiv 
-\frac{1}{2}(E_{LL} - E_{VV}- E_{PL} + E_{PV})
\end{equation}
\begin{equation}
 \mu \equiv \mu_P+
\frac{1}{2}(E_{LL} + E_{VV}+ E_{LV}- E_{PL}
- E_{PV})-\log 2
\end{equation}
The constant $\log 2$ appears in the definition of $\mu$
in order to render the simplified
expression (13) in a simpler form.

The above expression for ${\cal G}$
is studied in Sect. III for different
values of the various parameters and
the corresponding bulk phase-diagrams are obtained. For the study
of protein profiles, one can further simplify this
expression.  First, for small values of the order parameters,
i.e., relatively close to the critical point of demixing of the lipid and 
for small protein concentrations,
it is legitimate to expand 
the free energy of mixing up to
order ${\cal O}(\eta^4)$
and ${\cal O}(\phi^2)$. 
In addition, since
typical proteins occupy a much larger area then lipids,
the area ratio is in the range of
$\alpha  \sim 50 - 100$, and the protein entropy terms 
(of order  $1/\alpha$) can be  neglected in (7). 
The validity of the latter ($\alpha \rightarrow \infty$)
approximation will be reexamined
in Sect. III\cite{note1}.
With these simplifications, 
the approximated free energy density
can be written as
\begin{equation}
 {\cal F}^0/{T}=
-J \eta^2 +\frac{1}{12} \eta^4 +L \phi^2
+ \lambda \eta \phi +\frac{1}{2} \eta^2 \phi
\end{equation}
where the simplified thermodynamic potential using (8) is given
by ${\cal G}^0/{T}={\cal F}^0/{T}- \mu \phi -\mu_{\eta} \eta$.
The free energy density (13) needs some further discussion. 
Coexistence between dense ($\eta >0$) and dilute
regions ($\eta<0$) requires  that
$J>0$ and a
positive fourth-order term $\eta^4$ is needed to stabilize
the free energy. 
The protein itself is assumed not to be close to any phase transition.
Hence $L>0$ and no higher order terms in $\phi$ are needed.
We include in the expansion only the two lowest coupling terms 
between the protein and lipid concentrations.
The  first
is the  bilinear coupling 
$\eta\phi$ and has an enthalpic origin. 
It reflects the overall preference of the protein to more condensed
($\lambda<0$) or more dilute ($\lambda>0$) regions of the lipid monolayer. 
The second coupling is the symmetric $\eta^2\phi$ term, 
which is invariant under 
$\eta \rightarrow -\eta$ transformation
and provides the driving force for the localization
of proteins at the LE-LC interface. 
In our mean-field model, taking into account only pair
interactions, this coupling
has a purely entropic origin. More generally, it can 
also include interaction terms of higher-order in a virial expansion. 
Finally, the higher-order
coupling terms $\eta^2 \phi^2$ and $\eta^4 \phi$ are not  considered here 
since we try to investigate the most simple 
and yet non-trivial type of coupling.
A similar free energy
coupling has been introduced in the context of
polymer adsorption at liquid-liquid interfaces,
where in analogy the polymer adsorbs preferentially
at the interface from the bulk solution\cite{Pincus}.

For the case where the proteins in the 
monolayer are in equilibrium with a solution of proteins in
the aqueous subphase, the protein chemical potential  $\mu$
corresponds to the free energy
of adsorbing proteins  from the subphase into the monolayer
and depends on the concentration of proteins
in the subphase; this is  discussed in Sect. VIII.
Since we consider an insoluble (Langmuir) monolayer, similar
considerations do not apply to the chemical potential 
$\mu_{\eta}$  of the 
lipid order parameter $\eta$.
In fact, $\mu_{\eta}$ will be uniquely determined by the requirement of 
coexistence between dense and dilute lipid regions.
For
proteins which are {\it insoluble} in the subphase, 
the chemical potential $\mu$ acts as a Lagrange 
multiplier fixing the total amount of protein in the
monolayer, which is a conserved quantity in  this situation.

In the LE/LC two-phase region, obtained for $J>0$, one finds 
experimentally\cite{Mohwald}  domains of 
typically circular shape of LC phase
 immersed in a background of LE phase.
 Since the domains are rather large ($\sim 10-100 \mu m$),
we neglect
the shape of the line boundary  between the LC and LE
regions and  assume variation of the lipid
concentration  only along one spatial direction (the $x$ direction) 
and translational  invariance along the 
perpendicular direction.
The free energy $\gamma$ per unit length of
this  line boundary (related to the line tension
$\tau$ of the interface as calculated in Sect. VII)
is given by 
\begin{equation}
\gamma = \int_{-\infty}^{\infty} {\cal I} \d x
\end{equation}
where the free energy density ${\cal I}$ includes contributions
associated with spatial variations of the concentrations.
Defining the  ``stiffness coefficients''
 $g_\phi$ and $g_\eta$ for the protein and lipid
 concentration profiles, respectively, the free energy density 
 ${\cal I}$ is given by
\begin{equation}
{\cal I}= {\cal G}/{T} +\frac{1}{2} g_{\phi}
\left( \frac{\d \phi}{\d x} \right)^2 +\frac{1}{2} g_{\eta}
\left( \frac{\d \eta}{\d x} \right)^2
\end{equation}
In the next section we study the bulk phase diagram based on the
thermodynamic potential (8).
In the subsequent sections we use the simplified
expression (13) and determine the
concentration profiles $\phi(x)$ and $\eta(x)$ 
by applying a variational principle to
 the free energy functional $\gamma$.

\section{The phase diagram}
The phase diagram as a function of the 
chemical potentials $\mu_{\eta}$ and $\mu$ can be obtained from the 
thermodynamic potential (8) by
minimizing ${\cal G}$ with respect to the order parameters $\eta$ and $\phi$
in the two-phase coexistence region\cite{note2}. 
The coexisting solutions,
denoted by $(\eta_1,\phi_1)$ and $(\eta_2, \phi_2)$,  are 
determined from the equations
\begin{equation}
\mu_{\eta}= 
\left. \frac{\partial {\cal F}}{\partial \eta} \right|_{\eta_1,\phi_1}=
\left. \frac{\partial {\cal F}}{\partial \eta} \right|_{\eta_2,\phi_2}=
\frac{{\cal F}(\eta_1,\phi_1)-{\cal F}(\eta_2,\phi_2)}{\eta_1-\eta_2}
\end{equation}
\begin{equation}
\mu+\log 2=
\left. \frac{\partial {\cal F}}{\partial \phi} \right|_{\eta_1,\phi_1}=
\left. \frac{\partial {\cal F}}{\partial \phi} \right|_{\eta_2,\phi_2}
\end{equation}
which  correspond to a common-tangent construction.
These equations can be easily solved numerically.
In order to estimate the role of the
protein-lipid area ratio, $\alpha$, and to compare the results with the
calculations presented in the next section based on the simplified
expression (13), where $\alpha \rightarrow \infty$,
 we restrict the numerical analysis to the 
values $J=1/10$ and $L=10$.
The small value of $J$ means that one is close to the critical point
of the lipid phase separation, and the expansion in powers of
$\eta$, leading to (13),
 is appropriate. The large value of $L$ means that the protein 
concentration is rather small everywhere and can be treated as a 
small perturbation. We  will need this assumption  for the analytic solution of 
the Euler-Lagrange 
 equations in Sect. IV. The parameter $\alpha$ will be scanned in 
a rather wide range. With this choice of $L$ and $J$, it is clear that
the simplified free energy expression (13) is asymptotically obtained
for $\alpha \rightarrow \infty$.

The protein concentrations in the coexisting dense and
dilute lipid regions scan a whole range of different values, 
depending on the values of the remaining parameters $\mu$ and 
$\lambda$, 
but are strictly bounded
below by $\sim \exp(-\alpha)$. In contrast,
the simplified  free energy expression (13) has solutions with 
non-zero and strictly zero protein concentrations,  because of the
$\alpha \rightarrow \infty$ limit. It therefore allows
for straightforward classification of the bulk protein ordering
into a phase  with finite protein concentration and a phase 
with no proteins at all.
We need a similar criterion
for the case of the full free energy expression (7) with
$\alpha$ finite, allowing us to 
distinguish in a categorical manner the presence of proteins 
from the absence of proteins, even in 
the inevitable presence of  an exponentially
small (in $\alpha$) protein concentration. We adopt the
 simple criterion which consists of calculating the
Laplacian of the protein concentration in the parameter space $(\mu, 
\lambda)$,
\begin{equation}
 \frac{\partial^2 \phi_i}{\partial \mu^2}+
\frac{\partial^2 \phi_i}{\partial \lambda^2}
\end{equation}
in the two coexisting phases $\phi_1$ and $\phi_2$.
This scalar quantity shows a pronounced line of maxima in the parameter
space, separating two phases with small and large concentrations of
proteins. The position of this ridge is determined numerically
and defined  as the boundary between the two phases  rich and
devoid of proteins, respectively, for each solution $\phi_i$.
The result of this operation leads to three
distinct phase regions and is shown 
in Fig. 1 for the values $\alpha=10$, $50$, and $200$. 
Anticipating the definitions (28) and (30), we present the results
in terms of the rescaled variables $a \equiv \mu/(3J)$ and
$c \equiv \lambda /\sqrt{3J/2}$. 
The results obtained for $\alpha=\infty$ 
are denoted by solid lines. In the region denoted ``no proteins'' 
both protein concentrations $\phi_1$ and $\phi_2$
 are very small (exponentially in $-\alpha$); in the region
``semi-localized'' only one concentration is small
while the other is finite 
(distinguished by the criterion described above),
and in the region ``delocalized'' both phases have finite 
protein concentrations.

In the next section we will calculate
 the protein profile explicitly and,
in addition, obtain a ``localized'' phase.
This phase cannot be distinguished from the ``no protein'' phase
by just looking at the bulk free energy. In fact, in this phase
there is a finite protein concentration only at a finite
distance from the boundary between the LE and LC  regions.
As one can see from Fig. 1,  the phase boundary for $\alpha=50$ 
(long dashes) is 
already fairly close to the asymptotic boundary ($\alpha \rightarrow 
\infty$, solid line), so 
that neglecting the protein entropy is already
a good approximation for moderately  large
macromolecules.

\section{Euler-Lagrange equations}

In this section we calculate the protein concentration
profile based on the
free energy expression (15).
Minimization of the line free energy $\gamma$ (14) leads
to the Euler-Lagrange equations
(denoting $\d \phi/\d x$ by $\phi^\prime$, etc.)
\begin{equation}
\frac{\partial {\cal I}}{\partial \eta} - \frac{\d}{\d x} 
\frac{\partial {\cal I}}{\partial \eta^\prime}=0
\end{equation}
\begin{equation}
\frac{\partial {\cal I}}{\partial \phi} - \frac{\d}{\d x} 
\frac{\partial {\cal I}}{\partial \phi^\prime}=0
\end{equation}
Using the full free energy of
mixing (7), one obtains two coupled second-order  and non-linear 
differential equations
of the form
\[
-\mu_{\eta}-(2J+1) \eta+\lambda \phi +
\frac{1}{2} \log \left( \frac{1+\eta-\phi}{1-\eta-\phi} \right) =
g_{\eta}\frac{\d^2 \eta}{\d x^2}
\]
\[
-\mu -\log 2 + 2L\phi +\lambda \eta +\frac{1}{2}
\log \left( \frac{4(1-\phi)^2}{(1+\eta-\phi)(1-\eta-\phi)} \right)+
\frac{1}{\alpha} \log \left( \frac{\phi}{1-\phi} \right) =
g_{\phi}\frac{\d^2 \phi}{\d x^2}
\]
For the actual calculation of concentration profiles,
we will use the simplified free energy expression (13),
leading to the more compact expressions
\begin{equation}
-\mu_{\eta}-2J\eta+\frac{1}{3}\eta^3 +\lambda \phi +\eta \phi=
g_{\eta}\frac{\d^2 \eta}{\d x^2}
\end{equation}
\begin{equation}
-\mu + 2L\phi +\lambda \eta +\frac{1}{2}\eta^2 =
g_{\phi}\frac{\d^2 \phi}{\d x^2}
\end{equation}
These are the same equations that were considered by 
Halperin and Pincus in the context of polymer adsorption
at liquid-liquid interfaces\cite{Pincus}.

Instead of solving (21)-(22) numerically, we recall that
for large values of $L$ we can
treat the protein area fraction as a small parameter.
As a zeroth-order approximation,
we neglect the terms depending on 
$\phi$ in (21) and 
obtain as a solution the lipid order parameter profile
$\eta_0(x)$
in the absence of proteins. This profile is  then
inserted into (22), yielding the protein profile
$\phi(x)$.
The validity of this approach, namely solving
the  equation (21) while neglecting the coupling
between $\eta$ and $\phi$ and inserting the 
solution into equation (22), is critically examined
in Appendix B. There, it is found that this approximation
indeed corresponds to the first term
in an expansion, in which  the protein concentration
functions as the expansion  parameter and which therefore
is valid for small protein concentrations.

To proceed, setting $\phi=0$ in (21) leads to 
\begin{equation}
\eta_0(x)=\eta_{\infty} \tanh (x/\xi_{\eta})
\end{equation}
with the definitions
\begin{equation}
\eta_{\infty} \equiv \sqrt{6J}
\end{equation}
\begin{equation}
\xi_{\eta} \equiv \sqrt{g_{\eta}/J}
\end{equation}
This is the solution  of the usual 4th order 
Ginzburg-Landau free energy expansion and is strictly valid here
only for the pure lipid.
The lipid order parameter varies between
$+\eta_\infty$ for $x\rightarrow \infty$
and $-\eta_\infty$ for $x\rightarrow -\infty$,
 and its width is characterized by the correlation
length $\xi_\eta$. 
The chemical potential $\mu_{\eta}$ is zero in the approximation
employed above.
The origin is chosen as the symmetric point 
between the liquid condensed phase 
($x>0$)
and  the liquid expanded phase ($x<0$).
Defining a rescaled length $u\equiv x/\xi_{\eta}$ 
and a rescaled protein density
$\Phi(x) \equiv 4L \phi(x)/(\eta_{\infty})^2$,
the second differential equation (22) is reduced to 
\begin{equation}
\Phi(u)-h(u)=b^2 \frac{\d^2 \Phi(u)}{\d u^2}
\end{equation}
with the inhomogeneous term  $h(u)$  given by 
\begin{equation}
h(u) \equiv a- \tanh^2(u) -c \tanh(u)
\end{equation}
The remaining rescaled parameters are
\begin{equation}
a \equiv \frac{2 \mu}{(\eta_{\infty})^2}
=\frac{\mu}{3J}
\end{equation}
\begin{equation}
b^2 \equiv \frac{J g_{\phi}}{2L g_{\eta}}=
\frac{\xi_\phi^2}{ \xi_\eta^2}
\end{equation}
\begin{equation}
c \equiv \frac{2 \lambda }{\eta_{\infty}}=
\frac{\lambda}{\sqrt{3J/2}}
\end{equation}
The parameter $a \sim \mu$ is the rescaled chemical potential,
$b$ is the relative stiffness of the lipid
concentration profile compared to the protein 
concentration profile, and $c \sim \lambda$ measures the
preference of the proteins for the 
dense  ($c<0$)  or dilute ($c>0$) lipid domains.
The correlation length of the protein distribution 
is defined by $ \xi_\phi \equiv \sqrt{g_{\phi}/2L}$.

The general solution of the second order differential equation (26) 
can be written as
\begin{equation}
\Phi(u)=A \sinh(u/b) +B \cosh(u/b)+a -\Phi_1(u)/b -c \Phi_2(u)/b
\end{equation}
where the functions $\Phi_1(u)$ and $\Phi_2(u)$ are
given in Appendix A.
The constants $A$ and $B$ have to be determined in 
accord with the boundary conditions.

\section{Protein distribution}

\subsection{Solution for the case $b=0$}
It is instructive to treat first the limiting
case where the stiffness of the protein
distribution vanishes, i.e.,  $g_{\phi}=0$ and
$\xi_{\phi}=0$. Then, one has $b=0$ and the solution of
(26) is trivially given by $\Phi(u)=h(u)$. 
This leads to the protein distribution 
\begin{equation}
\Phi^0(u)  = \left\{ \begin{array}{lll}
        h(u) & \mbox{for }     & h(u) \geq 0  \\
        0    & \mbox{for }  & h(u) < 0 \\
                \end{array} \right.
\end{equation}
where the restriction to a finite range in $u$ follows
since $\Phi^0(u)$  has to be positive.
In fact, for $b=0$,
only for $h(u) \geq 0$ the protein distribution
is correctly described by the differential
equation (26); 
inspection of the free energy density ${\cal I}$ 
in the limit $\alpha \rightarrow \infty$
shows that the value of $\Phi(u)$ which minimizes
${\cal I}$ for $h(u)<0$
is given by $\Phi(u)=0$.
This failure of the variational methods used in deriving
(26) is due to the fact that one requires $\Phi(u)$
to be positive in the limit of very large proteins,
$\alpha \rightarrow \infty$.

Hereafter, we choose
$c\geq 0$ with no loss of generality, since 
the problem defined by (26) and (27) is symmetric 
under a simultaneous  inversion of $c$ and $u$
($c\rightarrow -c$ and $u \rightarrow -u$). 
Using the asymptotic behavior of $h(u)$,
\begin{equation}
h(u)    =\left\{ \begin{array}{lll}
         a-1+c & \mbox{for }     & u \rightarrow - \infty\\
         a-1-c& \mbox{for }     & u \rightarrow +\infty\\
                \end{array} \right.
\end{equation}
the following classification emerges:
(i)~For $a\leq 1-c$, the protein distribution 
vanishes both for positive and negative values of $u$ at a
sufficiently large but finite distance from the interface
(which is located at $u=0$); one actually obtains a
nonvanishing,
{\em localized} distribution of proteins 
provided that $h(u)>0$ for some range of $u$,
but this cannot be seen from the bulk behavior;
(ii)~for $1-c<a \leq 1+c$, the distribution is {\em semi-localized}
and vanishes only for sufficiently large positive values of $u$ and 
stays finite as $u \rightarrow - \infty$, and
(iii)~for $a>1+c$ the distribution is {\em delocalized} and
stays finite in both  limits $u \rightarrow \pm \infty$.
These three regimes are in accord with the phase diagram 
obtained in Sect. III and Fig. 1 for finite $\alpha$
and in the $\alpha \rightarrow \infty$ limit.

An additional observation can be made
for $c \leq 2$, where $h(u)$ has one maximum located at
\begin{equation}
u_{max}=-\tanh^{-1}(c/2)
\end{equation}
with a height
\begin{equation}
h(u_{max})=a+c^2/4
\end{equation}
(in the limit $c \rightarrow 2$ one obtains
$u_{max} \simeq \log(2-c)/2$). 
Consequently, for $c \leq 2$, the line defined by  $a=-c^2/4$ 
 marks the border between a fourth regime
where the protein distribution vanishes identically
(for $ a \leq -c^2/4$) and the regime where this distribution
is  non-zero (for a finite distance from the boundary between dense
and dilute lipid regions).
Fig. 2 summarizes these  borderlines in a phase diagram,
which is in fact valid also for $b \neq 0$, 
as will be discussed in the 
next subsection. The localized regime
is shaded in gray and ends at a special point $S$, at which 
the maximum of the protein distribution is at infinity; 
as pointed out before, there 
is an overall symmetry around the $a-$axis ($c=0$).

The effective correlation length 
$\xi_{eff}$ for the proteins in the localized regime
can be estimated from the curvature of 
$\Phi^0(u)$ at the maximum $u_{max}$, 
\begin{equation}
\xi_{eff}^{-2} \equiv -h''(u_{max})=2(1-c^2/4)^2
\end{equation}
This  length
diverges as one approaches the special
point $S$, where the distribution becomes indefinitely broad.

\subsection{Solution for $b >0\;\;\;\;
-\;\;\;\;$ general considerations}

On physical grounds, the solution for non-zero
$b$, i.e., for a finite stiffness of the protein 
distribution, has to coincide with the 
solution found for $b=0$ in the preceding section
very far from the interface located at $u=0$.
This leads to the general boundary condition
\begin{equation}
\Phi(u)=\Phi^0(u) \;\;\; \mbox{for} \;\;\;
u \rightarrow \pm \infty
\end{equation}
where $\Phi^0(u)$ is given by (32) and the general
solution $\Phi(u)$ is defined to be the concentration
profile which minimizes the free energy functional (14).
In the following, we discuss the properties of the
general solution $\Phi(u)$ separately for the four 
regions distinguished in Fig. 2.

i) In the delocalized case, the boundary conditions (37)
occurring at infinity together with the differential 
equation (26) valid for the entire ($-\infty, \infty)$ range
in $u$ are sufficient to determine the distribution 
$\Phi(u)$.

For the other cases, the boundary conditions (37) have to be 
supplemented by additional conditions at finite
values of $u$; the distribution $\Phi(u)$
is described by (26) only in a finite interval of $u$.

ii) In the case
where $h(u)<0$ for all $u$, it follows 
from the requirement $\Phi(u) \geq 0$ that
$\Phi(u)-h(u) >0$ and thus
all possible solutions of (26) have strictly positive curvature
as can be seen by looking at (26).
The boundary conditions (37), which imply
that $\Phi(u)=0$ as $u\rightarrow \pm \infty$,
 can not be satisfied for any
non-vanishing solution of (26). Consequently, the protein distribution
which minimizes the free energy is given  identically 
by $\Phi(u)=0$. This vanishing solution 
 was also found for $b=0$.

iii) When 
$h(u)$ is positive in some finite interval of $u$ but negative
for  $u \rightarrow \pm \infty$,
all solutions of (26) which are positive definite everywhere
 have  positive curvature for
$u \rightarrow \pm \infty$ and are
not  compatible with the
boundary conditions as given by (37).
This  merely reflects the fact that (26) describes
the distribution  $\Phi(u)$ only in the finite
interval $u_1 \leq u \leq u_2$, in  which
$\Phi(u)>0$.  The same  
was found to be true for $b=0$ in the last section.
From (37) in combination with (32),  $\Phi(u)$ has 
to vanish for $u \rightarrow \pm \infty$, 
and  can be positive  for finite $u$. 
As follows from minimizing the free energy functional
$\gamma$ (14), the solution $\Phi(u)$
has to be smooth everywhere  and thus fulfills
$\Phi(u)=\Phi'(u)=0$ at the two boundaries $u=u_1$ and $u=u_2$.

Now the following statements can be made:
a) There have  to be intervals of $u$ where 
$\Phi(u)$ has negative curvature
in order to fulfill the  boundary conditions $\Phi(u)=0$ at
$u=u_1$ and $u=u_2$;
b) close to the boundaries $u=u_1$ and $u=u_2$, 
the curvature has to be 
positive in order to fulfill $\Phi'(u)=0$ at $u=u_1$ and $u=u_2$;
c) consequently, the solution $\Phi(u)$ crosses $h(u)$ at 
two values of $u$ inside the region bounded by
$u=u_1$ and $u=u_2$, 
at which  the curvature of
$\Phi(u)$ vanishes; this can be seen from (26).
It follows that the boundaries $u_1$ and $u_2$ 
do not coincide, which means that the protein distribution
$\Phi(u)$ does not vanish identically.
We conclude that whenever the
distribution $\Phi^0(u)$ does not vanish for $b=0$, 
it is non-vanishing
for any $b \neq 0$. Note that it is actually
possible to construct a solution $\Phi(u)$ in accord with the 
boundary conditions at $u_1$ and $u_2$ since the 
general solution (31) has two adjustable parameters $A$ and $B$.

iv) For the semi-localized case, the boundary condition
(37) applies to the solution of (26)  
for $u \rightarrow - \infty$ only.
The protein distribution is non-zero in the $u$ interval
$(-\infty, u_2)$ and the boundary value $u_2$ satisfies
$\Phi(u_2)=\Phi'(u_2)=0$.

Putting together these arguments
for the different regimes, it follows that the phase diagram in Fig. 2 
is valid for general $b>0$.

\subsection{Boundary conditions}
In the following, we specify the boundary conditions 
for general $b$
for the three different  cases showing non-vanishing 
protein distributions:

\noindent
In the delocalized regime,
the boundary conditions obtained from (37), (33), and (32) are
\begin{equation}
\Phi(\pm \infty)=h(\pm \infty)= a-1 \mp c
\end{equation}
These boundary conditions 
 determine the coefficients $A$ and $B$
of the general solution (31).

\noindent
In the semi-localized regime, one has the conditions
\begin{equation}
\Phi(- \infty)=h(- \infty)= a-1 + c
\end{equation}
and
\begin{equation}
\Phi(u_2)=\Phi'(u_2)=0
\end{equation}
which determine the position of the boundary value, $u_2$,
and the coefficients $A$ and $B$.

\noindent
In the localized regime, one has 
\begin{equation}
\Phi(u_1)=\Phi'(u_1)=0
\end{equation}
\begin{equation}
\Phi(u_2)=\Phi'(u_2)=0
\end{equation}
Here, the boundary conditions 
 determine $u_1$, $u_2$, $A$, and $B$.
 In what follows,
we always assume that $u_1 \leq u_2$,
with no restrictions on the generality.

In the following, we present explicit protein profiles
$\Phi(u)$  for the limiting cases $b=0$ and $b=1$. The latter
value corresponds to the case where the correlation lengths 
of the lipid and protein concentration profiles are
equal, $\xi_{\eta}=\xi_{\phi}$.
Also, for $b=1$, 
the general solution of the protein profile as given in
Appendix A can be written in a simpler analytical form.

{\em Delocalized case:} for $b=1$, the coefficients are
determined to be $B=\pi/2-2$ and $A=c(1-\pi/2)$;
the protein distribution, given by (31), then reads
\begin{equation}
\Phi(u)=a-2 +2 \tan^{-1}[\tanh(u/2)](c \cosh u -\sinh u)
+\pi(\cosh u- c \sinh u)/2
\end{equation}
Using the equalities
\begin{equation}
\tan^{-1}[\tanh(u/2)]= \tan^{-1}[\e^u]-\pi/4=
\pi/4-\tan^{-1}[\e^{-u}]
\end{equation}
the protein distribution can be rewritten as
\begin{equation}
\Phi(u)=a-2 +\pi \e^u (1-c)/2 +2 \tan^{-1}[\e^u](c \cosh u -\sinh u)
\end{equation}
or
\begin{equation}
\Phi(u)=a-2 +\pi \e^{-u} (1+c)/2 -2 \tan^{-1}[\e^{-u}](c \cosh u -\sinh u)
\end{equation}
in accord with the limiting values $\Phi(u)=a-1 \pm c$ for
$u \rightarrow \mp \infty$.

{\em Semi-localized case:} For $b=1$,
the boundary condition at $u=-\infty$ leads to the
relation $A=2+B+c-\pi(1+c)/2$. The protein distribution
can be written as
\begin{equation}
\Phi(u)=a-2 +\e^u (B+2-c \pi/2) +2 \tan^{-1}[\e^u](c \cosh u -\sinh u)
\end{equation}
which indeed satisfies  the boundary condition as given by (39).
The coefficient $B$ and $u_2$ are in turn determined by
the second boundary condition (40).

{\em Localized case:} for general $b$,
the boundary conditions (41) and (42) 
can be cast in a more explicit form.
Defining
\begin{equation}
\cosh(u/b) \Phi(u) /b - \sinh(u/b) \Phi'(u) =B/b +\rho(u)
\end{equation}
with 
\begin{equation}
\rho(u) \equiv 
\cosh(u/b)(a/b-\Phi_1(u)-c \; \Phi_2(u))+
b \sinh(u/b)(\Phi'_1(u)+c \; \Phi'_2(u))
\end{equation}
and
\begin{equation}
\sinh(u/b) \Phi(u) /b - \cosh(u/b) \Phi'(u) =-A/b +\kappa(u)
\end{equation}
with 
\begin{equation}
\kappa(u) \equiv 
\sinh(u/b)(a/b-\Phi_1(u)-c \; \Phi_2(u))+
b \cosh(u/b)(\Phi'_1(u)+c \; \Phi'_2(u))
\end{equation}
leads to the equations
\begin{equation}
-B/b=\rho(u_1)=\rho(u_2)
\end{equation}
\begin{equation}
A/b=\kappa(u_1)=\kappa(u_2)
\end{equation}
Equations (48)-(51)
 have to be solved simultaneously in order to  determine
$u_1$, $u_2$, $A$, and $B$.
For the case $b=1$, the functions $\rho(u)$ and 
$\kappa(u)$ take the simpler form
\begin{equation}
\rho(u)=
(a-2)\cosh u +2 +\tanh u \sinh u +2c \tan^{-1}[\tanh(u/2)] -c \sinh u
\end{equation}
\begin{equation}
\kappa(u)=
(a-1)\sinh u +2 \tan^{-1}[\tanh(u/2)] +c (1-\cosh u)
\end{equation}

In the remainder of this section, we present protein profiles
calculated from the above equations for several values
of the three parameters $a$, $b$, and $c$.
Figure 3 shows protein distributions for four different values
of $a$ and for the two simple cases  $b=0$ (solid lines)
and $b=1$ (broken lines).
We set $c=0$, so the protein profiles are symmetric about
the LE/LC boundary located at $u=0$, where the lipid
concentration profile 
as given by (23) has an inflection point. 
For vanishing stiffness of the protein distribution ($b \rightarrow 0$), the profiles
have discontinuous slopes  for $a<1$ at the points where
the protein concentration vanishes;
the main effect of a non-vanishing stiffness parameter $b$ is to
eliminate these discontinuities, thereby flattening the entire  
concentration profile, as is clearly seen in Fig. 3.

Figure 4 shows asymmetric
protein distributions for four different values of $c$
on the transition line  between the localized and the semi-localized regimes,
defined by $a=1-c$.
Again, solid lines denote results for $b=0$ and broken lines
denote results for $b=1$.
As for the symmetric distributions shown in Fig. 3, a non-zero
stiffness parameter $b$ removes the discontinuity of
$\Phi'(u)$ at the boundary $u_2$ and flattens the concentration
profile.
As $c$ approaches the value $2$, the maximum of the distribution
moves progressively away from the LE/LC boundary located at $u=0$.
Also, the overall protein concentration rapidly decreases.
In the limit $c \rightarrow2$, the position of the maximum 
actually diverges logarithmically, as follows from  (34).

Figure 5 gives the localized protein distribution $\Phi(u)$ for
$c=0$ and $a=0.5$ for six different values of $b$, where 
$u_2$ and $B$ have to be determined numerically from
(42) applied to the general solution (31).
Interestingly enough, the boundary values $u_2=-u_1$ do not diverge 
as $b \rightarrow \infty$ but approach finite 
values $u_{1,2}= \mp 1.915$.
As the stiffness of the protein distribution increases,
the concentration is flattened and the area under
the curves decreases, but the profile does not spread out
indefinitely and stays localized.

\section{The protein excess }

The protein excess  is the total 
amount of adsorbed proteins. In the localized
regime, this quantity is defined as
\begin{equation}
\Gamma \equiv
\int_{-\infty}^{+\infty} \Phi(u) du=
\int_{u_1}^{u_2}  \Phi(u) du
\end{equation}
In the delocalized and the semi-localized regimes,
the quantity  $\Gamma$  as defined above diverges
since the protein distribution  approaches a constant
non-vanishing value as $u \rightarrow - \infty$ (for
the delocalized case the  same is also true as
$u \rightarrow\infty$).
One can still extract a meaningful quantity defined
by the excess amount of protein adsorbed 
by subtracting  the protein concentration at
$u= \pm \infty$, where $\Phi(\pm \infty)=a-1 \mp c$.
For $-2<c<2$ the protein distribution has one  maximum,
and we define the protein
excess as
\begin{equation}
\Gamma \equiv
\int_{-\infty}^{u_{max}} ( \Phi(u)-\Phi(-\infty)) du +
\int_{u_{max}}^{\infty} ( \Phi(u)-\Phi(\infty)) du
\end{equation}
where $u_{max}$ is the value of $u$ for which $\Phi(u)$ reaches its 
maximum.

\subsection{Protein excess for $b=0$}

The protein excess $\Gamma$
can be calculated for $b=0$
in closed form for all parameter values.
With  $\Phi^0(u)=h(u)=a-\tanh^2(u) -c \tanh(u)$,
the excess  can be written as 
\begin{equation}
\Gamma \equiv
\int_{u_1}^{u_2}  (a-\tanh^2(u) -c \tanh(u))du
\end{equation}
where the integration boundaries are given by
\begin{equation}
u_{1,2}=
\tanh^{-1}
 \left[-\frac{c}{2} \mp \sqrt{\frac{c^2}{4} +a} \; \right]
\end{equation}
For the symmetric case, $c=0$,
the boundaries $u_1$ and $u_2$ have the  values
\begin{equation}
u_{1,2}=
\mp \tanh^{-1} \sqrt{a}
\end{equation}
and  on the transition line between the
localized and the semi-localized regimes,
given by $a=1-c$,
one obtains $u_1=-\infty$ and
\begin{equation}
u_2=\tanh^{-1}(1-c)
\end{equation}
For general $a$ and $c$, the integral (58) yields
\begin{equation}
\Gamma=
\sqrt{c^2+4a}+(a-1+c)\tanh^{-1}\left[
\frac{\sqrt{c^2+4a}}{1+a} \right]
-c \tanh^{-1}\left[
\frac{\sqrt{c^2+4a}(2+c)}{2+2a+2c+c^2} \right]
\end{equation}
In the symmetric case, $c=0$,
this expression reduces to
\begin{equation}
\Gamma=
(a-1)\tanh^{-1}\left[
\frac{\sqrt{4a}}{1+a} \right]
+\sqrt{4a}
\end{equation}
and on  the localized to semi-localized
transition line,
$a=1-c$, it reduces to
\begin{equation}
\Gamma=
2-c-c\tanh^{-1}\left[
\frac{4-c^2}{4+c^2}\right]=2-c-c\log(2/c)
\end{equation}

Lines of constant $\Gamma$ for $b=0$ calculated 
from (62) are shown
as broken lines in Fig. 2.
Those lines can be helpful in interpreting
experimental findings when only the integrated protein
amount is known and not the entire profile.

\subsection{Protein excess for $b=1$}

For the symmetric case ($c=0$) the excess is given by
the closed-form expression
\begin{equation}
\Gamma=
2(a-1)u_2 + 2 \tanh u_2
\end{equation}
with the boundary value $u_2$ determined by
\begin{equation}
\tan^{-1}[\tanh(u_2/2)]=\frac{1-a}{2}\sinh u_2
\end{equation}
as follows from (53) and (55) and noting that $A=0$.

\noindent
For the localized to semi-localized transition line,
$a=1-c$, the excess is given by
\begin{equation}
\Gamma=
1-c \log(2) +\tanh u_2 -u_2 c -c \log(\cosh u_2)
\end{equation}
with the boundary value $u_2$ determined by 
\begin{equation}
2c+1=\tanh u_2+(1+c)(\cosh u_2-\sinh u_2)(\pi/2+
2 \tan^{-1}(\tanh[u_2/2])
\end{equation}
as follows from applying (42) to (47).

The protein excess $\Gamma$   for the
symmetric case $c=0$ is shown in Fig. 6(a)
as a function 
of $a$,
where the solid line denotes 
results for $b=0$ and the broken line for $b=1$.
These results correspond to the concentration profiles
plotted in Fig. 3 and are 
given by (63) and (65).
The protein excess for $b=1$ is smaller than for $b=0$,
which is also visible in Fig. 3. The overall flattening
of the distributions for non-zero $b$ causes the
area under the distribution to decrease.
For $a=1$, the
protein excess is given by $\Gamma=2$ for both values of $b$.
The same value holds for general $b$, as can be demonstrated
by numerical solutions of (56).
The boundary values $u_2$ given by (60) for $b=0$
and determined by (66) for $b=1$ are plotted in Fig. 7(a).

In Fig. 6(b) we show the protein excess 
on the localized/
semi-localized transition line, $a=1-c$,
as a function
of $c$,  as given by (64) and (67). 
As in the symmetric case, the protein excess 
$\Gamma$ decreases as $b$ becomes non-zero.
The boundary values $u_2$, obtained from  (61) and (68),
for $b=0$ and $b=1$, respectively,
are  plotted in Fig. 7(b).

\section{LE/LC line  tension in the presence of protein}

First we calculate the line tension 
of the liquid expanded-condensed interface in the
absence of proteins, denoted by $\tau_0$.
This energy per unit length
follows from the total free energy density
$\gamma$ as given by (14) after subtraction of the bulk
free energy density infinitely far from
the interface  and can be defined by
\begin{equation}
\frac{\tau_0}{2}= \int_0^{\infty} \d x \left\{
-J \eta^2_0(x)  +
\frac{1}{12} \eta^4_0(x) 
+\frac{1}{2} g_{\eta}
\left( \frac{\d \eta_0(x)}{\d x} \right)^2
+J \eta_{\infty}^2 
-\frac{1}{12} \eta_{\infty}^4
\right\}
\end{equation}
recalling that 
\begin{equation}
\eta_0(x)=\eta_{\infty} \tanh (x/\xi_{\eta})
\end{equation}
and
$\eta_{\infty}=\sqrt{6 J}$ and $\xi_{\eta}=\sqrt{g_{\eta} /J}$.
In writing (69) we used the symmetry around $x=0$.
The integral (69) is elementary and gives
the standard result 
\begin{equation}
\tau_0=
8 J^{3/2} g_{\eta}^{1/2}
\end{equation}
The total line tension is given by 
$\tau=\tau_0+\tau_{\phi}$.
The line tension contribution $\tau_{\phi}$ due to adsorbed proteins
in the localized phase region
can be written as 
\begin{equation}
\tau_{\phi}= \int_{-\infty}^{\infty} \d x \left\{
L \phi^2(x) 
- \mu \phi(x) +\lambda \eta_0(x) \phi(x)
+\frac{1}{2} \eta^2_0(x) \phi(x)
+\frac{1}{2} g_{\phi}
\left( \frac{\d \phi(x)}{\d x} \right)^2
\right\}
\end{equation}
For simplicity, we will restrict  ourselves to the limit $g_{\phi}=0$,
because the integration in  (72) can then be done in a closed form.
The line tension contribution $\tau_{\phi}$
can be expressed in reduced variables  as
\begin{equation}
\tau_{\phi}= \frac{\xi_{\eta} \eta_{\infty}^4}{8L}
\int_{-\infty}^{\infty} \d u \left\{
\frac{1}{2}  \Phi^2(u)  
-a \Phi(u) +c \; \tanh u \; \Phi(u) 
+ \tanh^2 u  \; \Phi(u) \right\}
\end{equation}
Using the solution found for $b=0$ (or, equivalently,
$\xi_{\phi}=0$),
$\Phi(u)=a - \tanh^2 u -c \tanh u$, the integral
can be solved for general $a$ and $c$.
Here, we only present the solution for the symmetric case,
$c=0$, which is given by
\begin{equation}
\tau_{\phi}= \frac{9 J^{3/2} g_{\eta}^{1/2}}{2 L}
\left\{ \sqrt{a} (1 - 5a/3) - (1-a)^2 \tanh^{-1}( \sqrt{a}) 
\right\}
\end{equation}
The limiting values are $\tau_{\phi}=0$ for $a=0$,
since in this case no proteins are adsorbed,
and $\tau_{\phi}=-3 J^{3/2} g_{\eta}^{1/2} /L$ 
for $a=1$, to be compared with $\tau_0=8 J^{3/2} g_{\eta}^{1/2}$.
This is the smallest  value
possible, for larger values of $a$ the line tension
contribution $\tau_{\phi}$ remains  constant.
The adsorption of proteins thus leads to a reduction
of the total line tension $\tau=\tau_0+ \tau_{\phi}$.
In principle, the total line tension $\tau$ 
can take negative values for
sufficiently large $a$ if
$L<3/8$, which  amounts to an instability of the LE-LC
interface,  possibly 
signaling a depression of the lipid phase transition.
Of course, in this limit the approximations used
in deriving (26) break down, since we assumed $L$ to be large
and the proteins being only a small perturbation
on the pure lipid phase transition.

\section{Protein profile in the subphase}

Up to now the coupled protein-lipid system was considered
as a pure two-dimensional system on the water/air interface
which is positioned at $z=0$. 
It is possible to evaluate the influence of a finite
solubility of the proteins in the subphase on the protein
distribution in the monolayer, and, in addition, to give a more
precise meaning to the protein parameters $\mu$ and $L$ used in (13).
The vertical protein concentration profile 
in the subphase can be calculated 
as a function of the distance $z$ from the monolayer,
For the calculation
of this
profile, which is denoted by $\phi_\perp (z)$,
we neglect any variation in the horizontal direction.
Assuming that the water is a good solvent for the protein
(and, therefore, that the aqueous 
protein solution is far from its
demixing curve), we can write the free energy per
unit area  on the surface as
\begin{equation}
\gamma_\perp=
\int_0^{\infty} \d z \left\{
\frac{1}{2} g^b_{\phi} 
\left(\frac{\d \phi_\perp(z)}{\d z}\right)^2
+L_b  \phi_\perp^2(z) -\mu_b  \phi_\perp(z) \right\}
+L_s \phi^2 - \mu_s \phi 
\end{equation}
where $\phi=\phi_\perp(0)$ is the protein concentration
at the surface  (or, equivalently,  in the monolayer). 
This expression is very similar to free energy functionals
studied in the context of wetting and other surface
phenomena\cite{Brezin}.
In analogy to the parameters used in (13) and (15),
$g^b_{\phi}$, $L_b$, and
$\mu_b$ are the protein parameters in the ``bulk'' subphase,
and $L_s$ and $\mu_s$ are the bare protein parameters at
the ``surface'' (or in the monolayer).
The chemical potential  $\mu_s$ measures the free energy
difference between a protein molecule  in the subphase and
in the monolayer, and it  contains contributions
due to van der Waals interactions of the protein with its
surrounding media as well as hydrophobic contributions 
coming from structural changes of the protein at the surface. It is
believed that proteins unfold their
hydrophobic parts when they are  inside 
a monolayer or even at the free air-water interface.
The energy
gained by such conformational transformations can be extremely high.

The Euler-Lagrange equation for the bulk density profile
takes the form 
\begin{equation}
2 L_b \phi_\perp(z) -\mu_b=
g^b_{\phi} \frac{\d^2 \phi_\perp(z)}{\d z^2}
\end{equation}
The bulk protein concentration infinitely far
from the monolayer is given from (76) by
\begin{equation}
\phi_b \equiv \phi_\perp (\infty) =\frac{\mu_b}{2 L_b}
\end{equation}
When the protein adsorbing on the surface is in contact
with a large bulk reservoir
of proteins,
the bulk concentration $\phi_b$ can be
regarded as a fixed
parameter and the chemical potential $\mu_b$ acts
as a Lagrange multiplier satisfying the relation $\mu_b
=2 L_b \phi_b$.

The solution of (76) compatible with the 
requirement $\phi_\perp(\infty)
=\phi_b$ is given by
\begin{equation}
\phi_\perp(z)=
(\phi -\phi_b) \e^{-z/\xi^b_{\phi}} +\phi_b
\end{equation}
where the correlation of the protein distribution
in the subphase is $\xi^b_{\phi}=\sqrt{g^b_{\phi}/2 L_b}$
and $\phi \equiv \phi_\perp(0)$ is the surface value.
The surface free energy, which is the total free
energy due to the presence of the monolayer at $z=0$,
 can be expressed as
\begin{equation}
\Delta \gamma_\perp=
\gamma_\perp -\int_0^{\infty} \d z \left\{
L_b  \phi_b^2  -\mu_b  \phi_b \right\}
\end{equation}
For the density profile given by (78), it takes  the form
\begin{equation}
\Delta \gamma_\perp=
(\phi - \phi_b)^2
\sqrt{L_b g^b_{\phi}/2} 
+L_s \phi^2
-\mu_s \phi 
\end{equation}
Minimizing this expression with respect to the
surface protein concentration $\phi$
leads to the value
\begin{equation}
\phi=
\frac{\mu_s + \phi_b \sqrt{2 L_b g^b_{\phi}}}
{2L_s+\sqrt{2 L_b  g^b_{\phi}}}
\end{equation}
Effectively, the presence of a finite concentration of
proteins  in the subphase can be
modeled by using the modified parameters
\begin{equation}
\mu=\mu_s + \phi_b \sqrt{2 L_b g^b_{\phi}}
\end{equation}
\begin{equation}
L=L_s+\sqrt{L_b  g^b_{\phi}/2}
\end{equation}
for the two-dimensional description of the protein distribution
in the monolayer, given in (13).
With these effective parameters, the $\phi$-dependent part
of the surface free energy can be 
rewritten (up to a constant) as
$
\Delta \gamma_\perp=
L \phi^2 - \mu \phi 
$.
For very large values of the protein adsorption
free energy $\mu_s$, as observed for proteins
which change their structure considerably as they
approach the air-water interface, 
and a rather small reservoir of proteins in the subphase,
most of the
proteins will be incorporated in the monolayer
leading to a depletion in the subphase, i.e.,
$\phi_b \approx 0$.
In this case the total
amount of protein in the  monolayer is a conserved
quantity and $\mu$ then acts as a Lagrange multiplier.
In the case of a large reservoir of proteins in the 
subphase, conservation of protein particles is taken care
of primarily by adjusting the bulk concentration
$\phi_b$.
In both cases, the parameter $\mu$, which appears in (13), 
can be tuned by changing the total amount of protein
added to the system.

\section{Effect of protein on surface pressure}

We discuss now how the parameters used in 
our calculation can be related to experimentally measurable
quantities, such as the lateral pressure $\Pi$.
This will be done for the simplified cases where there are 
either only proteins or only lipids at the water surface.
First, we calculate the lateral pressure in the limit of small
coverage with proteins or lipids, leading to a modified
ideal gas law; the correction to the limiting ideal gas behavior
gives information about the interactions.

In the case where no lipid is  present at
the water surface, one sets 
$\eta= \phi-1$ in (7) and the free energy
per lattice site  is
\begin{equation}
{\cal F}(\phi)/T= K \phi^2 +\{ \phi \log(\phi) 
+(1-\phi) \log(1-\phi)\}/ \alpha
\end{equation}
where $\alpha$ is the area ratio of the protein
and the underlying lipid/vacancy lattice.
$K \equiv L+\lambda-J-1/2$ 
is an effective  interaction parameter.
For the case where  only lipids are present one has to make the
replacements $K \rightarrow -4(J+1/2)$, $\phi \rightarrow \phi_L$ and 
$\alpha \rightarrow 1$.
The thermodynamic potential for a system covering
$N$ lattice sites is defined by
\begin{equation}
N {\cal G}= N {\cal F} -\mu T N \phi +\Pi N a^2
\end{equation}
with $a$ being the lattice constant of the underlying lattice
of vacancies or lipids.
Minimizing the potential with respect to the number of occupied
lattice sites $N$
and the protein density $\phi$ leads to
\begin{equation}
\mu=  \left. \frac{\d {\cal F}(\phi)/T}{\d \phi} \right|_{\phi=\phi_{eq}}
\end{equation}
\begin{equation}
-{\cal F}(\phi_{eq})+ \mu T \phi_{eq}=
\Pi a^2= \phi^2 \left. \frac{\d {\cal F}(\phi) /\phi}
        {\d \phi} \right|_{\phi=\phi_{eq}}
    = T K \phi_{eq}^2 -T \log(1-\phi_{eq})/\alpha
\end{equation}
Expanding the logarithm, one finds the behavior valid for
small surface pressures 
\begin{equation}
\Pi A = T +A^{-1} a^2 \alpha T (1/2 +K \alpha )
\simeq  T +\Pi  a^2 \alpha (1/2 +K \alpha )
\end{equation}
where $A=a^2 \alpha /\phi_{eq}$ is the surface area available
per protein (or lipid if one makes the replacement $\alpha=1$). 
The first term in (88) corresponds to the ideal 
gas behavior, the second term is an enthalpic and
entropic  correction from which the
effective interaction term $K$ can be deduced, if the area of a
protein, $a^2 \alpha$, and the protein-lipid area ratio $\alpha$
are  known. 

In order to estimate the critical interaction strengths,
it is useful  to define a new order parameter
$\theta\equiv 2 \phi-1$ for which the free energy expression (84)
can be expanded around $\theta=0$ and then reads (to  fourth order
in $\theta$ and neglecting terms linear in $\theta$)
\begin{equation}
{\cal F}/T=(\frac{K}{4}+\frac{1}{2 \alpha}) \theta^2 + 
\frac{\theta^4}{ 12 \alpha}
\end{equation}
from which the critical point of demixing is deduced to be
$K^*=-2/\alpha$. Note that for the case of lipids one recovers
the $\phi$-independent part of (13).
From (89) one sees
that for interaction
parameters $K>K^*/4$ the sign  of the correction in (88) is 
positive, vanishes for $K=K^*/4$ and actually becomes negative
as the interaction approaches the critical value. 
Measurements of $\Pi A$ as a function of $\Pi$ for the
hydrophobic polypeptide {\em cyclosporin} A  in the 
relevant temperature range indeed
showed positive slopes\cite{Wiedmann},
indicative of an interaction 
parameter $K$ far above  the critical value.
The sign of the parameter $\lambda$ indicates the preference
for the protein to enter dense ($\lambda<0$) or dilute
lipid regions ($\lambda>0$); experiments indicate that this parameter
is close to zero, so that $L$ is larger than zero. 
Neglecting higher-order terms in $\phi$ in
the free energy expression (13) thus seems justified,
assuming that {\em cyclosporin} A is a typical protein.

Next we show how the effective parameter $K$
can be related to properties of adsorbed
layer of proteins or lipids close to their 
critical points.
From (89),
the thermodynamic potential is given by
\begin{equation}
N {\cal G}=NT (\frac{K}{4}+\frac{1}{2 \alpha})
 \theta^2 +\frac{N T}{12 \alpha } \theta^4
- NT \mu  \theta + N \Pi a^2
\end{equation}
Above the critical point of demixing, defined by
the critical interaction strength $ K^*=-2/\alpha$,
one can neglect the fourth-order term and obtains
upon variation with respect to 
$N$ and $\theta$ a  relation between $K$ and the
equilibrium pressure $\Pi_{eq}$ and equilibrium
coverage $\theta_{eq}$ given by
\begin{equation}
\frac{K}{4}+\frac{1}{2 \alpha}
 =  \frac{a^2}{\theta_{eq}^2}
\left( \frac{\Pi_{eq}}{T} -\frac{\Pi^*}{T^*} \right)
\end{equation}
where
$\Pi^*$ and $T^*$ are the pressure and the temperature
at the critical point, thus material constants of the protein
(or the lipid).

Below the critical point of demixing and in the 
coexistence region one has to keep the fourth
order term and obtains the analogous relation
\begin{equation}
\left(\frac{K}{4}+\frac{1}{2 \alpha} \right)^2=
\frac{a^2}{3 \alpha}
\left( \frac{\Pi_{coex}}{T} -\frac{\Pi^*}{T^*} \right)
\end{equation}
where $\Pi_{coex}$ denotes the pressure
at coexistence  at a given temperature $T$.

For fitting experimental data to the above expression
it is important to note that the interaction parameters
$J$, $L$, and $\lambda$  as defined by (9-11) depend on the 
temperature.

\section{Discussion}

We studied a simple model which explains possible
aggregation
of proteins or other large macromolecules
at  the boundary  between coexisting
liquid condensed and liquid expanded domains of lipids.
Such a preferential adsorption of proteins has been 
observed experimentally\cite{Schwinn}.
Based on the general phase diagram, shown in Fig. 2,
obtained in 
the limit of proteins with large areas compared
with lipids ($\alpha 
\rightarrow \infty$), we predict a
transition from 
protein distributions localized at the LE/LC boundary
to semi-localized and delocalized distributions,
for which the protein concentrations remain finite 
in the coexisting lipid phases.
Such a transition can be observed by either changing
the total amount of adsorbed proteins (corresponding
to a change in $a$), or by changing
the temperature (influencing  the parameter
$c$).
We also calculated various experimentally accessible quantities,
such as the protein excess $\Gamma$ and the line tension $\tau$.
The line tension is predicted to decrease upon
adsorption of proteins.

The mechanism leading to the preferential adsorption of proteins at
the one-dimensional boundary line between LE and LC phases
is due to a  competition of the different
contributions to the entropy of mixing of the three components:
proteins, lipids, and vacancies. 
We recall that vacancies are artificially introduced just to
allow the Langmuir monolayer to be compressible.
Our model
assumes that the protein actually penetrates into the monolayer.
A partial intrusion is also possible and can be described by the model,
if the proteins take up at least some area at  the air-water interface.
Other mechanisms
based on long-ranged interactions such as electrostatic
forces are also important and
could  lead to similar results.

The affinity of the proteins to the LE/LC boundary can also
originate from other enthalpic reasons:
If the protein itself has amphiphilic properties 
with respect to the density of the surrounding medium,
i.e., if one moiety  of the protein favors a denser environment 
while the other moiety favors a more dilute environment,
it  would be driven into the interface between
the LE and LC phases. However,
such an amphiphilic  property of the proteins
seems to be unlikely, and,  if present, too
weak to produce the effects observed in experiments.

Finally, we mention that similar effects should be observable for 
freely suspended multicomponent membranes
which show phase separation into coexisting domains with 
different lipid compositions\cite{Bloom,Thewalt,Wu}.
Here, integral membrane proteins should be either dissolved in one 
of the domains, depending on the enthalpic preference, or, if
this preference is very weak, enriched and localized  at the one-dimensional
boundary line between the domains.

\acknowledgments
We would like to thank A. Goudot, H. M\"ohwald,
E. Sackmann, M. Schick, and T. Schwinn
for helpful discussions.
DA acknowledges partial support from the German Israel
Foundation (GIF) under grant No. I-0197 and
the US-Israel Binational Science Foundation (BSF)
under grant No. 94-00291.
RN acknowledges support from the Minerva Foundation,
receipt of a NATO stipend administered by the DAAD, and
partial support by the National Science Foundation under
Grant No. DMR-9220733.

\appendix

\section{Solution of differential equation}
Here we derive  the solution of the differential equation 
\begin{equation} 
\Phi(u)-h(u)=b^2 \frac{d^2 \Phi(u)}{du^2} 
\end{equation} 
with the inhomogeneous term  $h(u)$ given by 
\begin{equation} 
h(u) \equiv a- \tanh^2 u -c \tanh u 
\end{equation} 
Denoting by $\Phi_A$ and $\Phi_B$ two independent 
solutions of the homogeneous differential equation 
$\Phi(u)=b^2 \Phi''(u)$, 
the particular solution of the inhomogeneous 
differential equation is formally given by 
\begin{equation}
\Phi_P(u)=
-\frac{1}{b^2} \int_0^u dw \,\,h(w)
\frac{\Phi_A(w) \Phi_B(u)-\Phi_A(u) \Phi_B(w)}
     {\Phi_A(w) \Phi'_B(w)-\Phi'_A(w) \Phi_B(w)}
\end{equation}
Choosing $\Phi_A(u)=A \sinh(u/b)$,  $\Phi_B(u)=B \cosh(u/b)$,
and defining the particular solution as
\begin{equation}
\Phi_P(u)=a-\frac{1}{b} \Phi_1(u) -\frac{c}{b} \Phi_2(u)
\end{equation}
the integrals to be solved are 
\begin{equation}
\Phi_1(u) \equiv
\int_0^u dw \tanh^2(w) \sinh\left(\frac{w-u}{b}\right)
\end{equation}
\begin{equation}
\Phi_2(u) \equiv
\int_0^u dw \tanh(w) \;\; \sinh\left(\frac{w-u}{b}\right)
\end{equation}

\noindent
The integration is straightforward and yields 
\begin{eqnarray}
\Phi_1(u) &=&
b-b\cosh(u/b) + \sinh(u/b) \nonumber \\ &&
+\frac{1}{4b}\e^{-u/b}
\left(\Psi\left[\frac{1}{2}+\frac{1}{4b}\right]-
\Psi\left[\frac{1}{4b}\right]  \right)  +
\frac{1}{4b}\e^{u/b}
\left(\Psi\left[\frac{1}{2}-\frac{1}{4b}\right]-
\Psi\left[-\frac{1}{4b}\right]  \right)   \nonumber \\
&& + 2b F\left[2;-\frac{1}{2b};1-\frac{1}{2b};-\e^{2u}\right]+
2b F\left[2;\frac{1}{2b};1+\frac{1}{2b};-\e^{2u}\right]
\nonumber \\ && -
2b F\left[1;-\frac{1}{2b};1-\frac{1}{2b};-\e^{2u}\right]-
2b F\left[1;\frac{1}{2b};1+\frac{1}{2b};-\e^{2u}\right]
\end{eqnarray}
\begin{eqnarray}
\Phi_2(u) &=&
b-b\cosh(u/b) \nonumber \\ && +
\frac{1}{4}\e^{-u/b}
\left(\Psi\left[\frac{1}{2}+\frac{1}{4b}\right]-
\Psi\left[\frac{1}{4b}\right]  \right) -
\frac{1}{4}\e^{u/b}
\left(\Psi\left[\frac{1}{2}-\frac{1}{4b}\right]-
\Psi\left[-\frac{1}{4b}\right]  \right)   \nonumber \\ &&
-b F\left[1;-\frac{1}{2b};1-\frac{1}{2b};-\e^{2u}\right]-
b F\left[1;\frac{1}{2b};1+\frac{1}{2b};-\e^{2u}\right]
\end{eqnarray}
where $\Psi[z]$ denotes the {\it digamma function}  and 
$F[\alpha;\beta;\gamma;z]$ denotes the 
{\it hypergeometric function}\cite{Abramowitz}.
These special functions are defined by
\begin{equation}
\Psi[z]= \frac{\d [\log ( \Gamma[z])]}
             {\d z}
\end{equation}
with the gamma function defined as usual as
\begin{equation}
\Gamma[z]=
\int_0^{\infty} t^{z-1} \e^{-t} \d z
\end{equation}
and
\begin{equation}
F[\alpha;\beta;\gamma;z]=
\frac{ \Gamma[\gamma]}{\Gamma[\beta] \Gamma[\gamma-\beta]}
\int_0^1 t^{\beta-1} (1-t)^{\gamma-\beta -1}
(1- tz)^{-\alpha} \d t
\end{equation}
For the special case $b=1$, the above expressions simplify
and can be expressed as 
\begin{equation}
\Phi_1(u) = 2-2\cosh(u) + 2 \tan^{-1}[\tanh(u/2)] \sinh(u)
\end{equation}
\begin{equation}
\Phi_2(u) = \sinh(u)-2 \tan^{-1}[\tanh(u/2)] \cosh(u)
\end{equation}

\noindent
The general solution of the differential 
equation  is given by
\begin{equation}
\Phi(u)=A \sinh(u/b) +B \cosh(u/b)+a -\Phi_1(u)/b -c \;
\Phi_2(u)/b
\end{equation}
where the constants $A$ and $B$ are determined from
 the boundary conditions (see text).

\section{Low protein concentration expansion}

Here we discuss the validity of the approximations 
leading  from the Euler-Lagrange equations
(21) and (22) to the differential equation (26).
Namely, the use of the solution $\eta_0(x)$ in (22)
which was obtained  by
neglecting  the two terms 
$\lambda \phi$ and $\eta \phi$ in (21).
The solution $\eta_0(x)$ of the simplified differential
equation obtained by setting $\phi=0$ in (21) can be regarded
as a zeroth-order approximation to the full solution in 
an expansion in powers of $\phi(x)$. 
The validity of
this approximation 
can be estimated by reconsidering
the differential equation (21)  and substituting
for $\phi$ the solution $\phi(x)$ which was found 
initially by neglecting the coupling terms between
$\eta$ and $\phi$ in (21).

Consider first the
second coupling $\eta \phi$
between $\phi$ and $\eta$ in (21). This term
is unimportant as long as 
$\phi \ll J$. This is a reasonable assumption given that
the protein concentration is  small and one
is not too close to the critical point of 
the liquid-expanded liquid-condensed lipid
transition. This term
will not be considered any further.

In order to estimate the effect  of the other term 
which was neglected,
$\lambda \phi$, we define
\begin{equation}
\eta(x) \equiv \eta_0(x) + \delta \eta(x)
\end{equation}
with $ \eta_0(x)$ given by (23) and 
$ \eta(x)$ denoting the exact solution of (21).
Since  $\eta_0(x)$  solves  equation
(21) without the terms proportional to $\phi$,  the
differential equation for $\delta \eta (x)$ neglecting
terms of ${\cal O}(\delta \eta^2, \delta \eta  \phi)$ is given by
\begin{equation}
\delta \eta(x) (-2J +\eta^2_0(x))+\phi(x)(\lambda+\eta_0(x))=
g_{\eta} \delta \eta''(x)
\end{equation}
From the differential equation (21), one sees that
the correction we are estimating here
is important only for
\begin{equation}
\lambda \phi  \gg |-2J \eta_0  +\eta^3_0/3| \simeq
|-2J \eta_0|
\end{equation}
The last  step follows since $\eta_0(x)$ has to be much smaller 
than unity for the inequality to hold. This can only be true in the
close vicinity to the interface between dense and dilute
lipid regions, i.e., for $x \approx 0$.
Consequently, the correction $\delta \eta (x)$ is only important
around $x=0$.
Then, the terms proportional to $\eta_0(x)$
can be neglected and the differential equation (B2) 
simplifies to
\begin{equation}
-2J \delta \eta(x) +\lambda \phi(x) = g_{\eta} \delta \eta''(x)
\end{equation}
Replacing $\phi(x)$ by its value at the origin,
$\phi(0)$,
the solution of (B4) is  formally written as
\begin{equation}
\delta \eta(x)  =
\frac{\lambda \phi(0)}{2J} + C \sin(\sqrt{2} x/\xi_{\eta})
                          + D \cos(\sqrt{2} x/\xi_{\eta})
\end{equation}
In order for the correction $\delta \eta(x)$ to vanish
outside the region of interest centered around
$x \approx 0$, both coefficients $C$ and $D$ have to be of the
order as the constant $\lambda \phi(0)/2 J$.
The magnitude of the correction is thus given by
\begin{equation}
\delta \eta \simeq \frac{\lambda \phi(0)}{J} \sim
\frac{ c \phi(0)}{\eta_{\infty}}
\end{equation}
Note that $c$ is a parameter of order unity (or smaller) in 
the localized protein region (see Fig. 2).
Thus, the correction $\delta \eta$ enters in the 
calculation of the protein distribution $\phi(x)$ as a higher order 
contribution in terms of the ratio $\phi(0)/\eta_{\infty}$,
which is a small parameter. Neglecting this correction
is a controlled approximation corresponding to keeping
only the first order in a general expansion in terms of
$\phi(0)/\eta_{\infty}$, the ratio of the protein concentration and
the lipid concentration difference.

\begin{figure}
\caption{Bulk phase diagram for
$L=10$, $J=1/10$, and for $\alpha=10$ (short dashes), $50$ 
(long dashes), and $200$ (long-short dashes) as 
a function of the rescaled chemical potential 
parameter $a$ and the interaction parameter $c$.
The solid lines denote the phase boundaries for the 
limiting case $\alpha \rightarrow \infty$.
In the delocalized-phase region the protein concentration
in the dense and dilute lipid regions is finite;
in the so-called ``semi-localized'' region  only
the dilute lipid region contains proteins (for negative
values of $c$ only the dense lipid region contains proteins),
and in the region denoted by ``no proteins'' the protein
concentration is very  small ( $\sim \exp[- \alpha]$)  in both
coexisting lipid regions. 
In part of the  ``no protein'' region,
the solution of the Euler-Lagrange equations gives
a new localized protein distribution in the neighborhood
of the LE/LC boundary,
see Fig. 2.}

\caption{
Phase diagram for general $b$, valid in the
limit $\alpha \rightarrow \infty$. The shaded area denotes
the localized regime, in which the
 protein distribution is localized at
the boundary between dense and dilute lipid regions.
The special point $S$ is the limiting
localized case where  the maximum of the protein
distribution is infinitely far away from the
interface between the liquid condensed and liquid expanded
phases. The broken lines denote lines of constant 
excess protein $\Gamma$ calculated for the special case $b=0$.
The phase diagram is symmetric with respect to the
$a$ axis ($c=0$ line).}

\caption{
Protein distributions for the symmetric case defined by $c=0$
for different values of $a$; solid lines denote $b=0$ and broken
lines denote $b=1$.}

\caption{ Asymmetric
protein distributions at the boundary between the localized
case and the semi-localized case, defined by $a=1-c$;
solid lines denote $b=0$ and broken lines denote $b=1$.
The left boundary $u_1$ is located at $u_1=- \infty$ 
for all values of $b$.}

\caption{
Protein distribution $\Phi(u)$ for $c=0$ and $a=0.5$ for
the following values of $b$ (from top to bottom):
$b=0$, $0.2$, $0.6$, $1$, $1.6$, and $2.8$.
The limiting value of $u_2$ for $b \rightarrow \infty$ is
given by $u_2 = 1.915$.}
 
\caption{
(a) Protein excess $\Gamma$ for the symmetric case
$c=0$; the solid line denotes  $b=0$ and the broken line denotes 
$b=1$. At $a=1$ the excess is $\Gamma=2$
independently of the value of $b$.
(b)  Protein excess $\Gamma$ on the
localized/semi-localized transition line, defined by $a=1-c$.}

\caption{
(a) Boundary value $u_2$ for the symmetric case
$c=0$ for $b=0$ (solid line) as given
by (60) and for $b=1$ (broken line) as determined by (66);
note that here $u_1=-u_2$.
(b) Boundary value $u_2$ for the
localized to semi-localized transition line, defined by $a=1-c$,
for $b=0$ (solid line) as given
by (61) and for $b=1$ (broken line) as determined by (68);
note that here $u_1=-\infty$.}

\end{figure}


\begin{references}
\bibitem{note1}
It turns out that taking the limit $\alpha \rightarrow \infty$
 corresponds to neglecting the entropy contribution from
the proteins while maintaining $\phi \geq 0$, i.e., keeping a 
positive protein concentration.
\bibitem{note2}
Two-phase coexistence is observed on a whole line in the 
chemical-potential plane $(\mu_\eta, \mu)$; three-phase
coexistence, which occurs for small or negative values of $L$
and large values of $\phi$,
is not considered here.
\end{references}
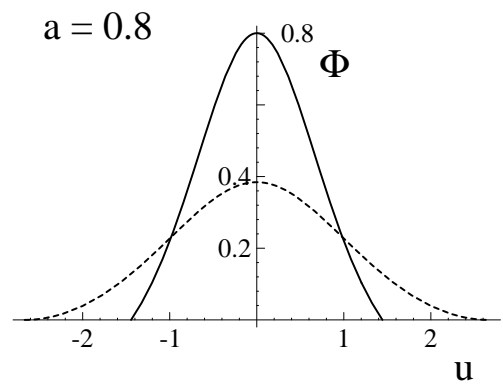
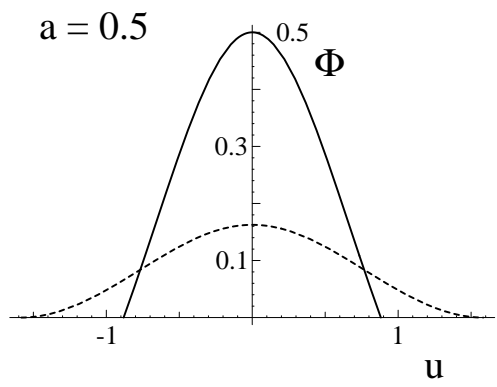
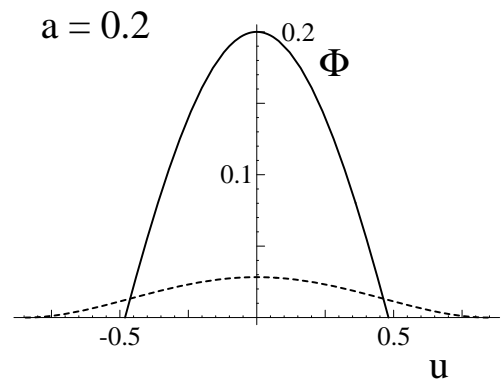
\end{document}